# Graphene Based Opt-Thermoelectric Tweezers

Xianyou Wang, Yunqi Yuan, Xi Xie, Yuquan zhang*, Changjun Min*, and Xiaocong Yuan*

Nanophotonics Research Center, Shenzhen Key Laboratory of Micro-Scale Optical Information Technology, Shenzhen University, Shenzhen, China, 518060

*Corresponding author: yqzhang@szu.edu.cn, cjmin@szu.edu.cn and xcyuan@szu.edu.cn

**ABSTRACT:** Since the discovery of graphene, its excellent physical properties has greatly improved the performance of many optoelectronic devices and brought important technological revolution to optical research and application. Here, we introduce graphene into the field of optical tweezers technology and demonstrate a new thermoelectric optical tweezers technology based on graphene. This technology can not only reduce the incident light energy by 2 orders of magnitude (compared with traditional optical tweezers), but also bring new advantages such as much broader working bandwidth and larger working area than the thermoelectric optical tweezers based on gold film widely studied before. Compared with gold film, graphene has more novel characteristics like high thermal conductivity, high uniformity and easy process. Thus, we found even monolayer graphene can achieve stable trapping for particles in a broad band, and the performance is enhanced with more graphene layers. Furthermore, structured graphene patterns can be easily generated to holographically trap multiple particles as desired shapes. This work verifies the great application potential of two-dimensional materials in optical tweezers technology, and it will promote more promising applications in cell trapping, tapping or concentration of biomolecules, microfluidics and biosensors.

Introduction

Optical tweezers, as a powerful scientific tool, have been broadly used to trap and manipulate small samples including bacteria, cells, quantum dots, and various metallic/dielectric particles (*1-4*). It has been widely applications in single-molecule biophysics, neuroscience and other bio-molecular applications(*5-7*). In traditional optical trapping regimes, a strong gradient optical force is necessary to overcome the scattering force, where a relatively high incident laser energy is often required. This, however, is inevitable to produce strong thermal effects in the focus area, resulting in two major problems: one is the possibility of thermal damage to samples, specifically, the common optical power density of conventional optical tweezers is in $10-10^3$ mW μm$^{-2}$, for which fragile targeted objects such as biological cells could be possibly damaged during the optical manipulation(*8, 9*); Second, it causes stronger Brownian motion, which makes the particle trapping accuracy worse especially for nano-particles.

In order to solve the above two problems, opto-thermoelectric tweezers (OTET) were proposed in 2018(*10*), which exploit the thermophoretic matter migration of charged ions under a light-induced temperature field to form a local electric field to trap the charged particles. The OTET greatly reduce the laser power required in traditional optical tweezers down to two orders, and provide a new platform for manipulating particles with a wide range of materials, sizes, and shapes(*11*) in low power. It has been successfully used in colloidal assembly(*12*), colloidal printing(*11*), and cell trapping(*13, 14*). OTET usually use a 5nm gold film and a 532nm laser as the trapping source(*10, 12, 15*). However, due to the limitation of absorption spectrum and fabrication of the 5nm-thick gold film, OTET are greatly limited in the selection of incident wavelength and film uniformity. The resonant absorption wavelength of gold film is about 500 nm, but such short-wavelength light is not very safe to biological samples such as cells(*9, 16*) or make quench fluorescence(*8*). The 5nm-thick gold film is usually not a uniform membrane, but a film with random nano island(*10, 17*), which could affect the heat transfer or form random hotspots to cause thermal damage to biological samples.

In order to solve these problems caused by gold film, we propose a new thermoelectric optical tweezers technology based on graphene. As a well-known two-dimensional material with advantages of wide optical absorption spectrum(*18*), high surface uniformity, good biocompatibility(*19, 20*), and high transparency, graphene has been widely used in different optical field(*21, 22*). In this paper we first introduce the principle of graphene using in thermoelectric optical tweezers, and then systematically studied the trapping capability of OTET based on graphene both theoretically and experimentally. Compared to the gold film, we find that graphene has more advantages in OTET. For example, even mono-layer graphene shows a much wider working wavelength range and gets better trapping stability than the 5nm gold film in 1064nm laser. The number of graphene layers can be adjusted to further enhance the trapping stability, and generate a large-scale trapping area. Furthermore, we demonstrate that graphene can be easily processed to various micro-structure patterns by the simple method of laser direct writing, which is very useful for holographic particle trapping to form arbitrary patterns. Although similar functions can be realized in metallic micro-/nano-structures based plasmonic tweezers(*23-25*), however, complicated and expensive electron beam lithography or focused ion beam (FIB/EBL) processing is generally required. Based on our proposed simple method, we have realized structured graphene patterns, and verified that particles can be trapped and arranged into a variety of designed patterns, and can even be combined with spatial light modulator (SLM) to dynamically manipulate the particle patterns. This work has established a thermoelectric optical tweezers system based on graphene for the first time, which offers a new idea for the combination of two-dimensional materials and optical tweezers technology, and also provides a new platform for the researches of on-chip holographic optical tweezers, biological detection chip, microfluidic and others.

## Results

As shown in Fig.1a, the schematic diagram of the graphene-based thermoelectric optical tweezers contains glass substrate, mono-layer graphene, sample polystyrene (PS) particles (blue ball), and surfactant (orange and green balls). The surfactant chosen here is cetyltrimethylammonium chloride (CTAC)(*10*). The CTAC molecules provide the macro-cations (CTAC micelles, orange balls in Fig.1a), while their counter-ions, Cl$^-$, serve as the anions (green balls in Fig.1a). The adsorption of CTAC cations on the PS particle surface also unifies the surface positive charges in the solution. As the incident laser is focused onto the graphene layer, the light absorption of graphene creates a localized non-uniform temperature field in the focal region, and then both the cations and anions undergo thermophoresis along the temperature gradient of the field and build an electric field pointing from cold to hot, which provides a strong thermoelectric force to trap the positive PS particle at the central thermal hotspot.

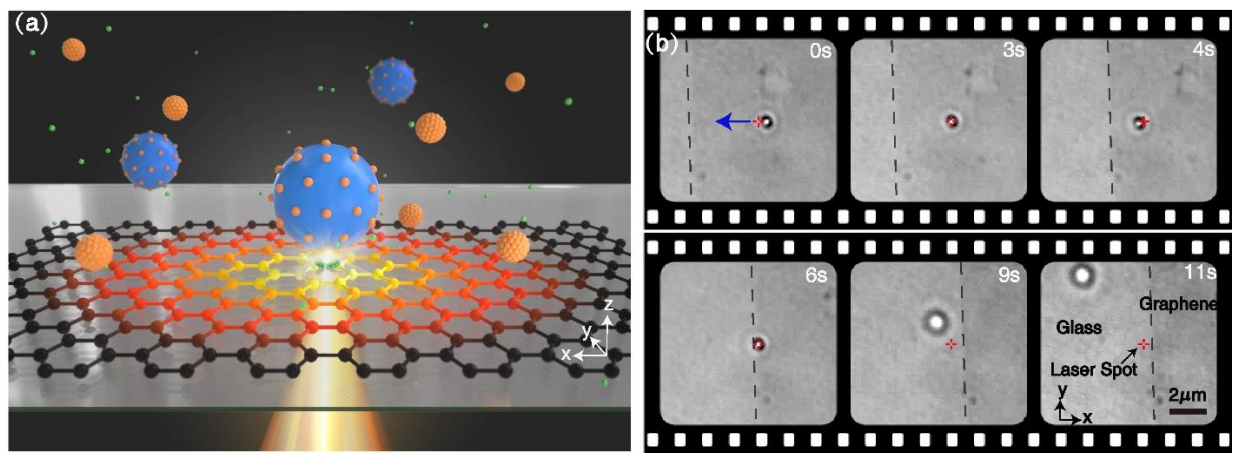

Figure 1. Schematics of graphene based opt-thermoelectric tweezers. (a) A schematic diagram of a thermoelectric optical tweezers based on a graphene substrate. The green particles are Cl$^-$ ions, the orange particles are CTAC cations, and the blue particles are the PS particles. (b) Successive images of a 500nm-diameter PS particle stably trapped and dragged to the left by a 1mW 1064nm laser on the graphene surface. The blue arrow indicates the drag direction, the dashed line is the graphene edge (the left area is glass and the right area is graphene), and the red cross is the center of laser spot.

Based on the above principles, we successfully trapped particles on the graphene substrate at quite a low laser power, as shown in Fig.1b. The trapping laser has 1064 nm wavelength with a power of ~1mW, and the corresponding optical system is shown in Fig5b and described in the Methods section. The process for the trapped particle dragging towards the edge of graphene was record in Fig.1b and in Supplementary Movie S1, where the red cross is the laser spot, the black dotted line indicates the edge of graphene and the blue arrow is the drag direction. Clearly, the particle can be stably trapped and moved on the graphene surface, but it escaped when the particle was dragged out of the edge of the graphene. Because there is no thermal gradient out of the graphene, and the optical gradient force at such low power is completely unable to trap the particle, so that the particle left the center of the laser spot and escaped. In our experiments, the trapped particle is 500nm-diameter PS particle, the monolayer graphene substrate is grown by chemical vapor deposition (CVD), and the concentration of surfactant CTAC is 3mmol/L (Based on the optimization shown in Supplementary Fig.S1). All details about the material preparation procedures are described in the Methods section. From these experimental results, we can find that graphene is fully suitable trapping particles at 1064 nm which is commonly used in optical tweezers(9).

To understand the experimental observations, we performed numerical analysis on the trapping forces and the potential well. On graphene surface two kinds of forces mainly contribute to the trap: optical force from the light focus and thermo-electric force induced by the thermal gradient of graphene. In order to distinguish the different effects of the two forces on the particles, a finite element modelling (FEM) model based on COMSOL Multiphysics was developed (details described in the Method section). On the basis of the theoretical model, for a 1064 nm laser beam with an average incident power of 1.5 mW, the thermal gradient distribution around a 500 nm-diameter PS particle located at the focal spot was calculated. In Fig2.a, the large temperature gradient is mainly concentrated around the particles with the maximum value of $1.22 \times 10^7$ K / m, where the red arrow represents the direction of the temperature gradient around the particle in the thermal field of the mono-layer graphene.

Based on the above thermal gradient simulation, the thermoelectric force on the particle as shown in Fig. 2b was calculated by the following Equations (26) 1 and 2.

$$E_T = \frac{k_B T \nabla T}{e} \left( \frac{\sum X_i n c_i S_{Ti}}{\sum X_i^2 n_i} \right) \quad (1)$$

$$TE_{force} = \int \sigma E_{T,\parallel} dA \quad (2)$$

where $i$ indicates the ionic species, $k_B$ is the Boltzmann constant, $T$ is the environmental temperature, $e$ is the elemental charge, and $X_i$, $c_i$, and $S_{Ti}$ are the charge number, the concentration, and the Soret coefficient of ionic species $i$, respectively. $E_{T,\parallel}$ is the tangential component of thermoelectric field $E_T$, $\sigma$ is the effective surface charge density and $A$ is the particle surface integral calculus. The normal component of thermoelectric field $E_{T,\perp}$ does not affect the movement of particles on the graphene surface (27). Therefore, the thermoelectric force mainly exists in the tangential plane (horizontal xy plane). Based all the electromagnetic field simulated by COMSOL, the above Equations (1-2) and Maxwell stress tensor(28), the calculated thermoelectric force and the optical force on the PS particle at different positions away from the center of optical axis(x = 0) is shown Fig. 2b. From their comparison, it can be found that both thermoelectric force and optical force point to the center of optical axis, but the maximum thermoelectric force reaches 0.41pN, which is about 18 times larger than that of the optical force (0.022pN). This comparison proves that the graphene-induced thermoelectric force greatly increases the trapping stability and reduces the laser power required. At the same time, due to the good thermal conductivity of graphene, the area of thermoelectric force is also very large (at x = 6um, there is still a large thermoelectric force (0.085pN)), which corresponds to a large trapping range in the experiment. In contrast, the traditional optical force shown in Fig. 2b only works in a small range of x < 0.5um.

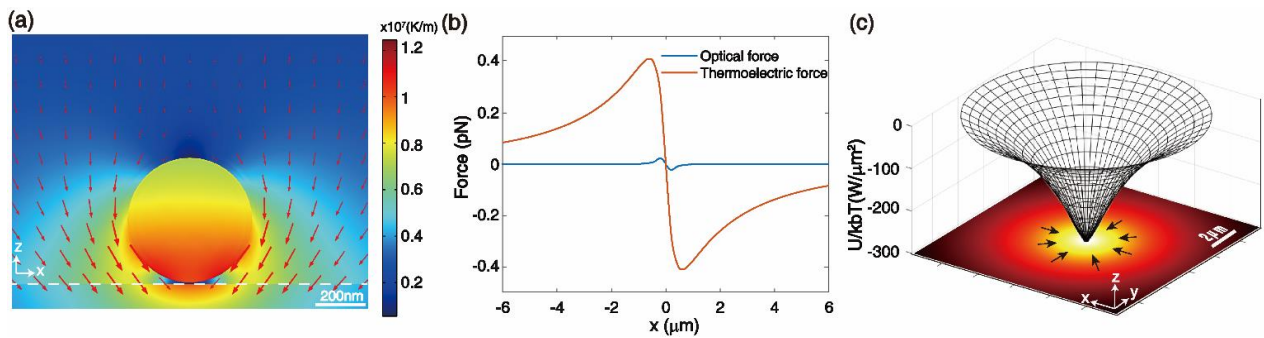

Figure 2. Numerical simulation of OTET based on graphene substrate. (a) The simulated distribution of the thermal gradient around a 500nm PS particle on the surface of mono-layer graphene at the focal point. The red arrow represents the direction of the temperature gradient around the particle, and the white dotted line represents the mono-layer graphene. (b) The thermoelectric and optical forces in *x* direction exerted on the PS particle at different positions around the center of optical axis ($x = 0$) on the graphene substrate. (c) The top figure shows the potential well distribution of the thermoelectric optical tweezers on the graphene substrate. The bottom figure shows the corresponding magnitude (background) and direction (black arrows) of the combined force (thermoelectric and optical force) distribution acting on the particle.

To prove the particle can be stably trapped and overcome the Brownian motion by the simultaneous action of thermoelectric force and optical force, we further calculated the potential well generated by the combined force with the formula $U = -\int_{\infty}^{r_0} \langle F(r) \rangle \, dr$, where $\langle F(r) \rangle$ is the optical force exerted on the particle located at $r$ and $r_0$ is the spot center (29). The calculated 3-dimensional potential well is displayed in the Fig. 2c, where the arrows indicate the direction of the combined force on the graphene substrate. It can be seen that the lowest point of the potential well is about $-300k_BT$, which is completely enough to overcome the Brownian motion (about $1\,k_BT$) and form a stable trapping. In addition, the action range of the potential well is large (more than 10 microns), which means that a large range of particles on the graphene substrate can be attracted and trapped to the center. The large trapping area is experimentally proved in Supplementary Fig.S2, which displays the dynamic process for a large range of particles trapping and assembling a superlattice with 15um diameter. This characteristic of graphene-based thermoelectric optical tweezers has potential applications in particle self-assembly(17)(30)and concentration of biomolecules or cells(31).

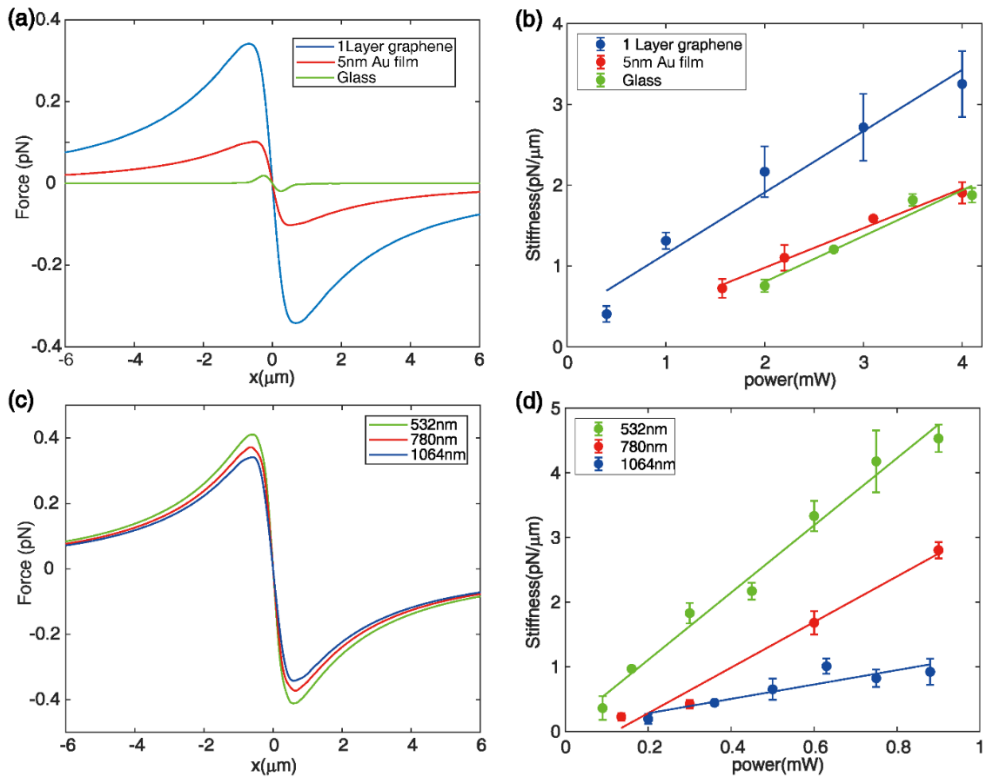

Figure 3. Trapping ability comparison among traditional optical tweezer, thermoelectric optical tweezers based on graphene and gold film substrate. (a) Theoretical simulation shows different total trapping forces on glass surface, 5nm-thick gold film and monolayer graphene under 1064nm laser with 1.5 mW incident power for the 500nm-diameter PS particles. (b) The experimentally measured trapping stiffness on the glass surface, 5nm-thick gold film and monolayer graphene under 1064nm laser with different laser powers. (c) Theoretical result of the total trapping force for 500nm PS particles on monolayer graphene at 532, 780 and 1064 nm laser wavelengths and 1.5 mW incident power. (d) The experimentally measured trapping stiffness for 500nm PS particle on monolayer graphene under different laser wavelengths and power.

In order to quantitatively compare the graphene-based thermoelectric optical tweezers with the previous gold-film-based one and the traditional optical tweezers without thermoelectric force, in Fig. 3a we theoretically calculated the total trapping force in the three cases with substrate of bare glass, 5nm-thick Au film and monolayer graphene, respectively. Obviously, for the case of traditional optical tweezers with only glass substrate, there is only optical force and no thermoelectric force, so the total force is relatively weak and the action range is small (only a few hundred nanometers limited by focal spot). Both cases with gold film and graphene substrates have optical force and thermoelectric force, in which thermoelectric force is dominant as Fig. 2b. However, as the graphene's absorption at 1064nm is much higher than that of gold film, its total trapping force is also much higher than the case of gold film (the corresponding thermal gradient distributions for the three cases are shown in Supplementary Fig. S3a-c). Meanwhile, as trapping stiffness can reflect the ability of optical tweezer in experiment, in Fig. 3b we experimentally measured the trapping stiffness on the three different substrates, and the method to calculate the trapping stiffness is introduced in the Methods section. In Fig. 3b, we can find the trapping stiffness increases almost linearly with the laser power, and the minimum trapping power of monolayer graphene is only about 0.2mW. In contrast, the minimum trapping power for pure optical trapping with glass substrate is more than 2mW. And the stiffness with the gold film is relatively close to the case of glass, indicating that in experiment the thermoelectric force induced by the gold film at 1064nm has much weaker effect than that at 532nm in previous reports (*10*), and is also weaker than the case of monolayer graphene. Thus, the experimental results are consistent with the theoretical prediction for the three cases. Additionally, the trajectories of trapped particles with 2mW incident power for the three cases are displayed in Supplementary Fig. S3d-f, which give the intuitive comparison of the trapping ability of the three cases. The trajectories of trapped particles for the monolayer graphene case show the smallest region of particle motion, in accordance with the results in Fig. 3b.

Next, to study the effect of broadband absorption characteristics of graphene on the OTET, we consider three different laser wavelengths (532nm, 780nm, 1064nm) focused on the monolayer graphene to trap single PS particle. The trapping forces at these three wavelengths are calculated theoretically as shown in Fig. 3c, which shows that the trapping force is the strongest at 532nm and the weakest at 1064nm. The corresponding thermal gradient distributions for the three wavelengths are shown in Supplementary Fig. S4a-c. The difference mainly comes from the different absorption of graphene at these wavelengths (*18*). In experiment, the trapping stiffness for the three laser wavelengths were also measured as shown in Fig.3d. We found the optical trapping stiffness increases almost linearly with the laser power, and the trapping stiffness is larger at shorter wavelength under the same laser power, agreeing with the theoretical prediction in Fig. 3c. Furthermore, the minimum incident laser power required to overcome Brownian motion to trap particles is 0.1mW under 532nm, which is far less than that of traditional optical tweezer of 2mW in Fig.3b. Additionally, the trajectories of trapped particles with incident power of 0.6mW for the three wavelengths are displayed in Supplementary Fig. S4d-f, where the area of trajectories of trapped particles increases with the wavelength, intuitively demonstrating the results in in Fig.3d.

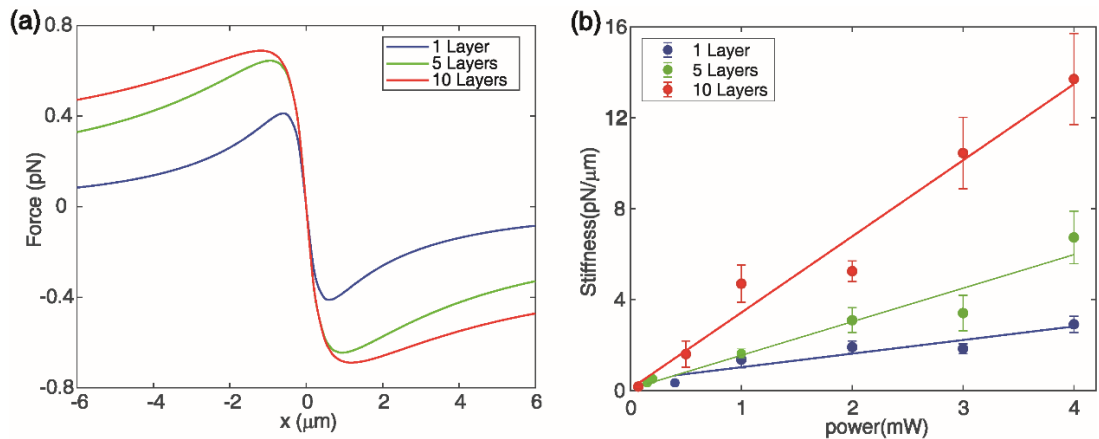

Figure 4. The increase of graphene layers enhances the trapping force of thermoelectric optical tweezers (a) Simulation results of the trapping force distribution for 500 nm PS particle with different graphene layers at 1064 nm. (b) Experiments results of the trapping stiffness for 500 nm PS particle with different graphene layers at 1064 nm.

Because of the linear electronic energy dispersion, the monolayer graphene is essentially transparent in the visible and near infrared band with an absorptivity of 2.3%(*32, 33*). Therefore, we can increase the light absorption rate of the graphene substrate by increasing the number of graphene layers, so as to enhance the thermoelectric force at the focal point. The simulation results show the thermal gradient fields with different number of graphene layers (Supplementary Fig.S5a-c), and the calculated trapping force is clearly enhanced as the graphene layers increases as shown in Fig. 4a. In experiment, we measured the trapping stiffness of single PS particle on 1, 5 and 10 layers of graphene substrates as a function of laser power as shown in Fig.4b. It is found that with the increase of graphene layers, the optical stiffness is enhanced significantly at the same laser power, which is consistent with the theoretical results. More importantly, the minimum trapping power required is as low as 0.07mW with the 10 layers graphene substrate, which is far less than the minimum trapping power of 2mW on monolayer graphene (Fig. 3b). Also, the trajectories of trapped particles with 1mW incident power shown in Supplementary Fig. Sd-f clearly verify the performance of the three different graphene layers.

## Multiple particles manipulation with graphene micro-structures

In addition to single particle trapping, we further studied the simultaneous trapping and pattern arrangement of multiple particles based on the graphene thermoelectric optical tweezers system. Usually, there are two main methods to trap multiple particles to form desired patterns: one is to generate multiple optical focal spots by holograms on phase modulation devices (e.g. SLM), such as holographic optical tweezers; the other is to generate multiple electromagnetic hot spots by micro-/nano-structures, such as structured plasmonic tweezers(34). For the first method, because of the ultra-high thermal conductivity of graphene(35), a single focused spot can form a wide thermal gradient field on the graphene surface, resulting in a large trapping area of particles as shown in Supplementary Fig. S2. Therefore, each trapping area of the multi focuses generated by SLM is easy to interfere with adjacent ones, making it hardly to form a clear pattern with multiple isolated particles on graphene surface. Thus, here we choose the second method, and propose to pattern the graphene film into discrete lattice, and then use a plane-wave light beam to generate discrete hotspots on the graphene lattices to trap multiple particles, as schematically shown in Fig.5a. We use a simple method of direct laser writing (details introduced in Methods section) to pattern graphene into the designed shapes, and then use a SLM as shown in Fig.5b to modulate the incident Gaussian beam into a uniform plane-wave beam and focus it onto the patterned graphene micro-structures. Figure 5b presents the optical setup for writing the graphene micro-structures and trapping the particles (details of the experimental setup described in Methods section). To demonstrate the performance of particle manipulation with the graphene micro-structures, we have successfully fabricated the graphene lattices as two designed patterns of letters S and Z as shown in Fig.5c and Fig.5e, respectively, and the corresponding experimental verification results are displayed in Fig.5d and Fig.5f, respectively.

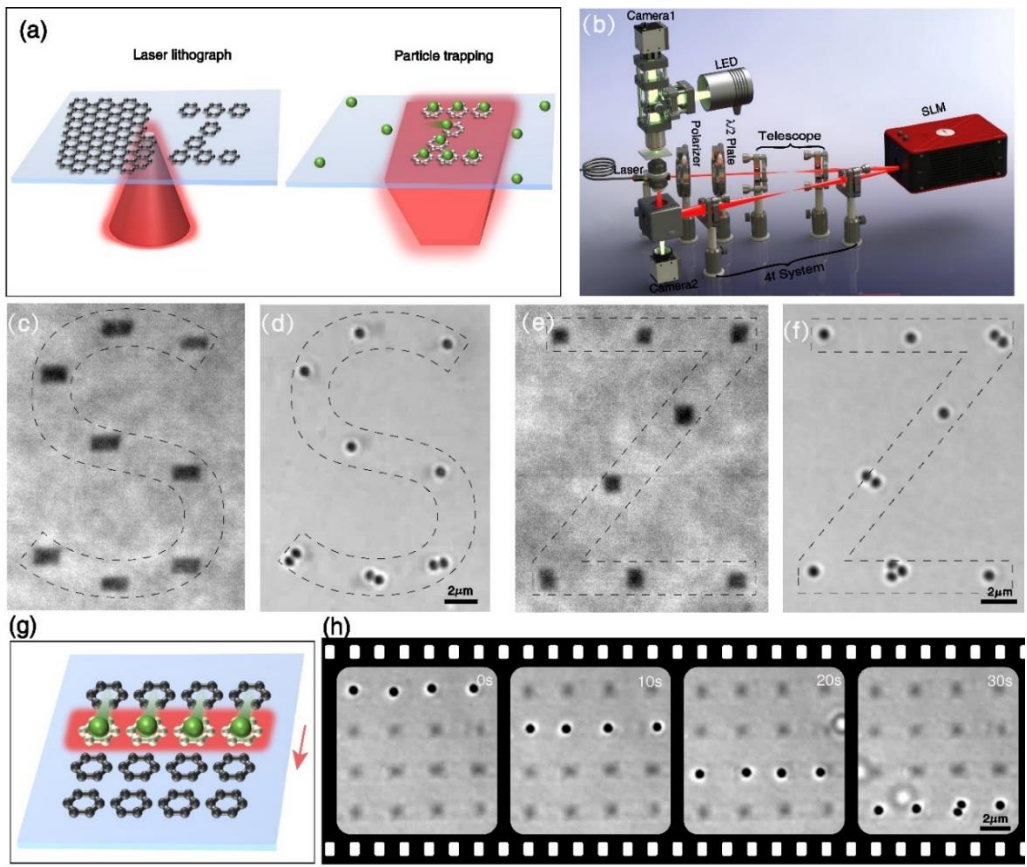

Figure 5. Patterning graphene thermoelectric optical tweezers for multiple particles trapping and movement. (a) Schematic diagram of patterning graphene thermoelectric optical tweezers. (b) Experimental optical configuration of the trapping system. (c) Direct laser writing method prints S-shaped graphene lattices, and (d) the corresponding experimental trapping effect for multiple

PS particles. (e) Direct laser writing method prints Z-shaped graphene lattices, and (f) the corresponding experimental trapping effect for multiple PS particles. (g) Schematic diagram of dynamic manipulation for multiple particles on graphene lattices, and (h) the corresponding experimental effect of dynamic movement of particle array.

In addition to arranging particles into static patterns, we further studied the dynamic regulation of particle patterns. With the combination of SLM's dynamic control and the graphene lattice, we can achieve dynamic manipulation for particles array as schematically shown in Fig.5g. In experiment, we use SLM to generate a line-shaped focused beam, and then dynamically move the PS particles array on the graphene lattice as shown in Fig.5h. The corresponding videos of trapping progress are displayed in Supplementary Movie S2. Compared with the similar function realized by structured plasmonic tweezers(23, 24), our method does not need complex and expensive micro-/nano-processing equipment such as EBL or FIB. Furthermore, this method can dynamically control the particles on each graphene lattice in combination with holographic light field, and can also be combined with microfluidic system for applications as high-throughput biological detection chip, drug screening and others.

## Conclusions

In this paper, we have successfully realized a new thermoelectric optical tweezers based on graphene substrate. Compared to the common-used gold film, we found even mono-layer graphene can extend the thermoelectric optical tweezers to a wider working wavelength range, such as 1064nm with less damage to cells or fluorescence. As the ultra-high thermal conductivity of graphene, we demonstrated this technology can achieve a large-area particle trapping or concentration, and the performance can be greatly enhanced by adding the number of graphene layers. We also verify that graphene can be easily processed to designed micro-structures by the laser direct writing method, and thus can be used for holographic particle trapping to form desired patterns, and can even be combined with SLM to dynamically manipulate the particle patterns.

It is worth noting that this technology can be further developed in many aspects. In this work, we only measured 3 different wavelengths that cannot cover the absorption spectrum of graphene, actually the working wavelength of the system can be extend to other bands. Besides, we only measured maximum 10 layers of graphene, more layers of graphene or other two-dimensional materials may further improve the effect. Moreover, this paper demonstrate graphene can apply in opto-thermoelectric tweezers, actually it can also suit for opto-thermophoretic tweezers with non-electrolyte solvent(36, 37). Additionally, the voltage can be applied to regulate absorption of the two-dimensional materials, resulting in electrically controlled dynamic optical tweezers. At the same time, the large-area and multiple fixed-point trapping with above graphene microstructures have application value in microfluidic chip system or particle sorting system. We believe this work has not only proposed a new optical technology or a powerful tool, but also demonstrated an example that the combination of two-dimensional materials and traditional optical technology could greatly improve the performance and promote many new applications.

## Methods

### Numerical Simulation

The finite elements method (COMSOL Multiphysics, 5.6) was used to simulate the electromagnetic field distribution and thermal field distribution around the single PS particle in the laser focused field at the substrate/solution interface. A two-dimensional model composed of a glass substrate, a graphene thin film and water was built. Electromagnetic Waves, Frequency (ewfd) Domain module and Heat Transfer (ht) in Solids and Fluids module were used for simulation. Laser was set to focus on the interface between the graphene (gold film) and the water medium, and the corresponding power was set to 1.5mW. Perfectly matched layers (PML) were placed around the simulation domain to prevent backscattering from the boundaries. The refractive index of graphene was chosen from Weber (38), the thermal conductivity, heat capacity at constant pressure and electrical conductivity of graphene were chosen as 5300 W/(m · K), $7^{10}$ J/(Kg · K)(39) and $10^8$ S/m(40), respectively. Transition Boundary Condition in ewfd module and Thin Layer in ht module were used to describe optical and thermal characteristic of graphene, respectively. The thickness of single layer graphene was set as 0.35nm. The thickness of gold film is 5nm. The diameter of PS particle is 500nm. The gap distance between PS particle and substrate is 5nm. The refractive index for water, glass, PS particle and gold film were 1.333, 1.5, 1.5983, 0.54386+2.2309i, respectively(41, 42). Room temperature (293.15K) was set at the outer boundary of the model. Electromagnetic Heating in Multiphysics was used for coupling heat generation. Maxwell stress tensor (MST) method(28) was used to calculate the optical force acting on the particle. The thermal properties of glass, PS particle, gold film and water were adapted from the COMSOL material library.

### Materials preparation

CVD-grown graphene films were purchased from 6Carbon technology company (Shenzhen). Different layer graphene films were transferred to the glass substrates by chemically detaching samples from the growth substrates using poly-(methyl methacrylate) PMMA thin film as a mediator(43). The 5nm-thickness Au films were prepared through thermal evaporation with speed 1A/S and followed by thermal annealing at 550 °C for 2 h. The PS particles with citrate-functionalized surfaces in different diameter were purchased from nanoComposix. The particle suspension was centrifuged for 5 min (3000g) and re-dispersed in CTAC solutions with the desired concentration.

### Optical configuration

The configuration is shown as Fig.5b. An inverted microscopic system (Olympus IX81) was established to achieve a stable trap for targeted particles. The linearly polarized 532nm laser beam (Changchun Leishi Photoelectric Technology Co., Ltd) expanded by a telescope system with two convex lenses. The expanded beam be modulated by SLM (Holowye PLUTO-VIS-0020) and project to the oil-immersed objective lens (Olympus ×100 NA=1.49) by 4f system. The gap between the objective lens and the glass substrate is filled with index-matching oil to satisfy the exciting condition. For dark field illumination, in-house built elements consisting of an objective (Olympus ×100 NA=0.9) and a high-speed CMOS camera (Pointgrey GS3-U3-23S6M-C) are used to capture the experimental process from the upside. Another CMOS camera (Pointgrey GS3-U3-23S6M-C) was used to capture the particle dynamics from down side. A custom-made LabVIEW program performed the video recording, laser control and synchronization of moving platform (ASI MS-2000) motion. For the single particle trapping the setup only need replace the SLM by a reflector and couple the 780nm (Newport VAMP TA-7613) and 1064nm (Changchun Leishi Photoelectric Technology Co., Ltd) laser source.

### Quantifying trapping stiffness

Measurements to quantify the trapping stiffness of trapped PS particle at varying conditions were performed on the previously described microscope system. The trap stiffness is then calculated by analyzing the recorded videos of the trapping process(44). To retrieve the motion data from the image sequences, an image registration technique, i.e. Phase-Correlation (PC) of the Fourier Transform method(45), is adopted to achieve sub-pixel motion resolution (46). The optical tweezer trapping force are determined using the Boltzmann distribution of thermally driven position fluctuations(47). As a fundamental method, only need a minimum of information: Boltzmann constant $K_B = 1.3807 * 10^{-23}$, the solution's temperature 293K and the mean square displacement retrieved in previous step. The whole image processing and stiffness calculation are performed by a custom-made MATLAB program.

### Direct laser printing

The optical setup of direct laser writing was based on the same system of optical tweezers with the objective lens (Olympus ×60 NA=0.9). The laser wavelength is 532nm and the laser power is 20mW. The number of graphene layers used for Multiple particle manipulation with graphene micro-structure is 10. The scanning step is 100nm. A custom-made LabVIEW program performed the graphics conversion and printing control.

## AUTHOR INFORMATION

### Author Contributions

X. Wang, and C. Min developed the concept presented in this work. X. Xie performed the FEM simulations. Y. Yuan and X. Wang performed the experiment system construction and experimental measurements. X. Wang and Y. Yuan performed the experiment data analysis. X. Wang, Y. Zhang and C. Min wrote the manuscript. C. Min and X. Yuan supervised all of the work and oversaw preparation of the manuscript. All authors discussed the results and commented on the manuscript.

### Notes

The authors declare no competing financial interest.

ACKNOWLEDGMENT

Guangdong Major Project of Basic and Applied Basic Research (2020B0301030009), National Natural Science Foundation of China (91750205, 61427819, 61490712, U1701661, and 61605117, 61975129), Leading Talents of Guangdong Province Program (00201505), Natural Science Foundation of Guangdong Province (2016A030312010, 2018A030310553, 2019TQ05X750), Science and Technology Innovation Commission of Shenzhen (KQTD20170330110444030, ZDSYS201703031605029, JCYJ20178181443338999, JCYJ20180305125418079), General Financial Grant from the China Postdoctoral Science Foundation (2017M612722). The authors would like to acknowledge the Photonics Center of Shenzhen University for technical support.